\documentclass[11pt, letterpaper]{article}

\usepackage[
    letterpaper,
    top=1in,
    bottom=1in,
    left=1in,
    right=1in,
    headheight=14pt
]{geometry}

\usepackage{iftex}
\usepackage{microtype}  % Improved typography (kerning, protrusion)
\usepackage[english]{babel}

\usepackage{amsmath}
\usepackage{amsthm}

\ifPDFTeX
    \usepackage[T1]{fontenc}
    \usepackage[utf8]{inputenc}
    \usepackage{stix2}
\else
    \usepackage{fontspec}
    \usepackage{unicode-math}
\fi

\usepackage{ragged2e}
\usepackage{setspace}
\usepackage{tikz}
\usetikzlibrary{positioning, arrows.meta, shapes.geometric, decorations.pathmorphing, calc}
\usepackage{graphicx}
\usepackage{float}
\usepackage[font=small, labelfont=bf, labelsep=period]{caption}


\usepackage{booktabs}
\usepackage{array}

\usepackage{fancyhdr}
\fancypagestyle{firstpage}{
    \fancyhf{}
    \fancyfoot[C]{\thepage}
    
}

\usepackage{titlesec}
\titleformat{\section}
    {\normalfont\large\bfseries}
    {\thesection.}{0.5em}{}
\titleformat{\subsection}
    {\normalfont\normalsize\bfseries}
    {\thesubsection}{0.5em}{}
\titleformat{\subsubsection}
    {\normalfont\normalsize\itshape}
    {\thesubsubsection}{0.5em}{}

\titlespacing*{\section}{0pt}{2.5ex plus 1ex minus .2ex}{1.5ex plus .2ex}
\titlespacing*{\subsection}{0pt}{2ex plus 1ex minus .2ex}{1ex plus .2ex}

\usepackage[numbers,sort&compress]{natbib}
\usepackage{hyperref}
\hypersetup{
    colorlinks=true,
    linkcolor=darkgray,
    citecolor=darkgray,
    urlcolor=darkgray,
    pdfauthor={Arthur Juliani},
    pdftitle={The Manifold of the Absolute: Religious Perennialism as Generative Inference},
    pdfsubject={Computational Theology, Variational Autoencoders, Religious Perennialism},
    pdfkeywords={perennialism, VAE, religious epistemology, manifold hypothesis}
}
\usepackage{xcolor}
\usepackage{colortbl}

\usepackage{thmtools}
\usepackage[framemethod=TikZ]{mdframed}

\definecolor{defcolor}{RGB}{240,248,255}      % Alice blue
\definecolor{propcolor}{RGB}{255,250,240}     % Floral white
\definecolor{remcolor}{RGB}{248,248,248}      % Near white

\mdfdefinestyle{defstyle}{
    backgroundcolor=defcolor,
    linewidth=0.5pt,
    linecolor=darkgray,
    roundcorner=3pt,
    innertopmargin=8pt,
    innerbottommargin=8pt,
    skipabove=12pt,
    skipbelow=8pt,
    nobreak=true
}

\mdfdefinestyle{propstyle}{
    backgroundcolor=propcolor,
    linewidth=0.5pt,
    linecolor=darkgray,
    roundcorner=3pt,
    innertopmargin=8pt,
    innerbottommargin=8pt,
    skipabove=12pt,
    skipbelow=8pt,
    nobreak=true
}

\mdfdefinestyle{remstyle}{
    backgroundcolor=remcolor,
    linewidth=0.5pt,
    linecolor=darkgray,
    roundcorner=3pt,
    innertopmargin=8pt,
    innerbottommargin=8pt,
    skipabove=12pt,
    skipbelow=8pt,
    nobreak=true
}

\declaretheoremstyle[
    headfont=\bfseries,
    notefont=\bfseries,
    bodyfont=\normalfont,
    mdframed={style=defstyle}
]{defbox}

\declaretheoremstyle[
    headfont=\bfseries,
    notefont=\bfseries,
    bodyfont=\normalfont,
    mdframed={style=propstyle}
]{propbox}

\declaretheoremstyle[
    headfont=\itshape,
    notefont=\itshape,
    bodyfont=\normalfont\small,
    mdframed={style=remstyle}
]{rembox}

\declaretheorem[style=defbox, name=Definition, numberwithin=section]{definition}

\declaretheorem[style=propbox, name=Claim, sibling=definition]{claim}
\declaretheorem[style=propbox, name=Assumption, sibling=definition]{assumption}

\renewenvironment{abstract}{%
    \begin{center}
    \begin{minipage}{0.9\textwidth}
    \small
    \noindent\textbf{Abstract.}
}{%
    \end{minipage}
    \end{center}
    \vspace{1em}
}

\date{}

\begin{document}

\thispagestyle{firstpage}
\centering
\vspace*{2em}
{\LARGE\bfseries The Manifold of the Absolute:\\[0.3em]
\Large Religious Perennialism as Generative Inference\\[1.5em]}
{\textbf{Arthur Juliani}\\[0.3em]
\normalsize Institute For Advanced Consciousness Studies\\[0.2em]
\small\texttt{arthur@advancedconsciousness.org}}
\vspace{2em}
\begin{abstract}
This paper formalizes religious epistemology through the mathematics of Variational Autoencoders. We model religious traditions as distinct generative mappings from a shared, low-dimensional latent space to the high-dimensional space of observable cultural forms, and define three competing generative configurations corresponding to exclusivism, universalism, and perennialism, alongside syncretism as direct mixing in observable space. Through abductive comparison, we argue that exclusivism cannot parsimoniously account for cross-traditional contemplative convergence, that syncretism fails because combining the outputs of distinct generative processes produces incoherent artifacts, and that universalism suffers from posterior collapse: stripping traditions to a common core discards the structural information necessary for inference. The perennialist configuration provides the best explanatory fit. Within this framework, strict orthodoxy emerges not as a cultural constraint but as a structural necessity: the contemplative practices that recover the latent source must be matched to the specific tradition whose forms they take as input. The unity of religions, if it exists, is real but inaccessible by shortcut: one must go deep rather than wide.
\end{abstract}

\justifying
\section{Introduction}

The philosophy of religion has long struggled to reconcile the multiplicity of religious forms with the singular claim to truth. How can mutually exclusive doctrines, from the Christian Trinity to Islamic \textit{tawhid} (divine oneness) to the Buddhist teaching of dependent origination (\textit{pratītyasamutpāda}), each claim privileged access to ultimate reality? Four dominant responses have emerged in the modern discourse: \textit{exclusivism}, which privileges one tradition as uniquely true; \textit{pluralism} (or universalism), which treats all traditions as culturally-conditioned approximations of a common core; \textit{syncretism}, which freely combines elements across traditions; and \textit{perennialism}, which posits a transcendent unity underlying distinct traditions, accessible only through committed practice within one.

The intuition that a single truth underlies diverse religious forms has deep roots. The Renaissance doctrine of \textit{prisca theologia}, articulated by Marsilio Ficino \cite{ficino1482}, held that a primordial divine wisdom was transmitted through a chain of ancient sages, including Zoroaster, Hermes Trismegistus, Pythagoras, and Plato, each clothing the same transcendent truth in the idiom of their own culture. The most systematic ancient articulation of this vision was Neoplatonism: Plotinus (204--270 CE) developed Plato's metaphysics into a comprehensive emanationist ontology in which all of reality flows from a single, absolutely simple source, ``the One'' (\textit{to Hen}), and the soul's task is to ascend back toward that source through contemplative purification \cite{plotinus_enneads}. Ficino was explicitly reviving Plotinus when he translated the \textit{Enneads} into Latin, and the Neoplatonic framework (emanation from unity, return through contemplation) became the dominant metaphysical grammar for perennialist thinking in the West \cite{omeara1993}. Agostino Steuco gave this intuition its enduring name in \textit{De perenni philosophia} (1540) \cite{steuco1540}, and Leibniz later adopted the term \textit{philosophia perennis} to describe his own project of reconciling rival philosophical systems by extracting their common truths \cite{schmitt1966}. The nineteenth century's comparative turn, exemplified by Schopenhauer's integration of Vedantic and Buddhist thought into Western metaphysics \cite{schopenhauer1818}, lent empirical substance to these earlier speculations.

In the twentieth century, Aldous Huxley's \textit{The Perennial Philosophy} \cite{huxley1945} brought the concept to a wide audience through an anthology of converging mystical testimonies. The ``Traditionalist'' school, founded by René Guénon \cite{guenon1925} and developed by Frithjof Schuon \cite{schuon1948} (later popularized by Huston Smith \cite{smith1976}; for a critical history, see Sedgwick \cite{sedgwick2004}), gave Perennialism its most rigorous metaphysical formulation. Although the terms ``Traditionalism'' and ``Perennialism'' are often used interchangeably, they diverge on a point we address in Section~\ref{sec:related}; this paper adopts the Perennialist view. Perennialism argues for a \textit{Sophia Perennis}, a transcendent unity underlying distinct exoteric forms, while simultaneously insisting that this unity can only be accessed \textit{through} strict adherence to a single orthodox tradition. This position has often appeared paradoxical: if the truth is one, why cannot one mix and match the paths that lead to it?

This paper addresses this apparent paradox by formalizing religious epistemology through the lens of Variational Autoencoders (VAEs) \cite{kingma2014} and conducting an abductive comparison of four competing models. We propose that the transmission of divine truth operates as a \textit{generative process}, where a simple, unified source (the latent variable $z$) is projected through tradition-specific decoders into richly elaborated observable cultural forms (the data $x$). We define three configurations of this generative model, corresponding to exclusivism, universalism, and perennialism, and analyze syncretism as direct combination of outputs in observable space. We argue that the Perennialist configuration provides the best explanatory account. Within this framework, the practitioner's task is to train an \textit{inference network}: an encoder ($q_\phi$), realized through contemplative and mystical practice, that can recover the latent source from the tradition's observable forms.

The power of this structural model lies in its capacity to articulate, rather than merely assert, the comparative advantages of the Perennialist position. We argue that syncretism fails under all three configurations: because valid religious data lie on non-linear manifolds, neither interpolation in observable space, interpolation in latent space, nor ad hoc decoder combination produces a valid hybrid inhabiting two existing manifolds (Section~\ref{sec:syncretism}). Exclusivism, meanwhile, cannot parsimoniously account for the well-documented convergences in cross-traditional contemplative reports (Section~\ref{sec:exclusivism}), while universalism suffers from \textit{posterior collapse}, since stripping traditions to their ``common core'' discards the structural information necessary for latent variable inference (Section~\ref{sec:universalism}). The Perennialist configuration alone preserves the encoder-decoder matching required to maximize the Evidence Lower Bound (ELBO), the formal criterion for successful inference (Section~\ref{sec:perennialism}).

We emphasize at the outset that the argument is \textit{abductive} and \textit{conditional}: it proceeds as an inference to the best explanation, and it depends on the empirical premise that cross-traditional contemplative convergence is genuine rather than merely apparent. What the formalization adds beyond informal Traditionalist arguments is a set of constraints that yield non-obvious consequences: manifold geometry articulates \textit{why} syncretism fails, not merely that it does; the Data Processing Inequality identifies universalism's impoverishment as structural, not incidental; and the ELBO decomposition reveals orthodoxy and self-transcendence as coupled terms in a single objective, not independent demands. The Traditionalist insistence on orthodoxy, often dismissed as mere conservative preference, may reflect a genuine structural constraint on any system seeking to recover transcendent meaning from immanent practice.

\section{Related Work}
\label{sec:related}

Our approach draws on three distinct literatures: the Traditionalist school in philosophy of religion, the contemporary debate on religious pluralism, and computational approaches to modeling belief systems.

\subsection{The Traditionalist School}

The Perennialist or Traditionalist school originated with René Guénon's \textit{Introduction to the Study of Hindu Doctrines} (1921) \cite{guenon1925} and reached its mature expression in Frithjof Schuon's \textit{The Transcendent Unity of Religions} (1948) \cite{schuon1948}. The central thesis holds that beneath the diversity of exoteric religious forms lies an esoteric unity accessible only through initiation into a living tradition. Subsequent contributors include Ananda Coomaraswamy \cite{coomaraswamy1977}, Titus Burckhardt \cite{burckhardt1967}, and Seyyed Hossein Nasr \cite{nasr1989}, who extended the framework to encompass Islamic spirituality and the philosophy of science. The Traditionalist school is not monolithic: Guénon's metaphysics is more rigidly hierarchical, tending to privilege Advaita Vedānta as the purest expression of the \textit{sophia perennis}, while Schuon's formulation treats traditions more symmetrically, as equally valid crystallizations of the same transcendent truth (see Sedgwick \cite{sedgwick2004} for a critical history). This internal diversity maps onto our framework: Guénon's position implies that some traditions' generative processes are more expressive than others, while Schuon's implies rough parity. Our perennialist configuration accommodates both readings (see Section~\ref{sec:discussion}).

A deeper divergence within the school bears directly on our argument. The Traditionalists proper, particularly Guénon, hold that all valid spiritual traditions descend from a single primordial Tradition (\textit{Sanātana Dharma} or \textit{Sophia Perennis}) transmitted through unbroken initiatic chains \cite{guenon1925}; the set of legitimate traditions is essentially closed, and no genuinely new tradition can arise \textit{de novo}. The broader Perennialist position, which this paper adopts, holds instead that transcendent truth can spontaneously manifest in new cultural and historical contexts, so that genuinely new traditions may emerge without requiring an unbroken line of transmission from a single originating revelation. The emergence of Sikhism from Guru Nanak's revelation (Section~\ref{sec:syncretism}) is more naturally accommodated by this open reading. The Perennialist position is both more parsimonious (it does not require positing untraceable transmission chains) and less exclusionary, since it does not privilege civilizations with documented ancient lineages over those without them.

\subsection{Religious Pluralism}

The Perennialist position must be distinguished from John Hick's ``pluralistic hypothesis'' \cite{hick1989}, which holds that all major religions are equally valid responses to a single transcendent ``Real.'' Hick's model more closely resembles what we term \textit{universalism}: the assumption that tradition-specific features are accidental rather than essential. Our analysis suggests that Hick's approach, while ethically commendable, suffers from an information-theoretic problem: stripping traditions to a common core discards the structural information necessary for inference (see Section~\ref{sec:universalism}).

Alvin Plantinga's ``exclusivism'' \cite{plantinga2000} and Gavin D'Costa's ``inclusivist'' position \cite{dcosta1986} represent alternative responses to pluralism. The exclusivist holds that salvation is available only through one tradition; the inclusivist allows that other traditions may participate in truth through their relation to the privileged tradition. Inclusivism is not a rival to our framework but a special case within it: it maps onto a hierarchical variant of the perennialist configuration in which all traditions share a latent source and possess distinct generative processes, but one tradition's process is taken to be maximally faithful to that source while others capture genuine but partial aspects of the same reality. The eliminative arguments of Sections~\ref{sec:exclusivism}--\ref{sec:universalism} apply equally to this inclusivist variant, since they require only a shared latent source with tradition-specific generative processes, not parity among them. The perennialist configuration thus subsumes both the egalitarian reading (Schuon) and the hierarchical inclusivist reading (D'Costa, Rahner's ``anonymous Christians,'' and much of Vatican~II theology); the formal structure is the same, and only the relative expressiveness of the generative processes differs.

\subsection{Variational Autoencoders and Multi-Modal Learning}

Within machine learning, the VAE framework \cite{kingma2014} has become central to understanding latent variable models. The phenomena of posterior collapse \cite{bowman2016} and the manifold hypothesis \cite{bengio2013} that we invoke are well-established in the deep learning literature. Particularly relevant is the multi-modal VAE literature, which directly studies the perennialist setup: shared latent spaces with modality-specific encoders and decoders. The MVAE \cite{wu2018}, MMVAE \cite{shi2019}, and MoPoE-VAE \cite{sutter2021} demonstrate that learning shared representations across modalities is feasible but reveals characteristic challenges: modality dominance (one modality's encoder captures the latent space more faithfully than others), posterior quality that varies across modalities, and a coherence-diversity tradeoff in which shared representations sacrifice modality-specific detail. These findings are broadly consistent with the perennialist picture: they confirm that a shared-latent-space, multiple-decoder architecture is a natural and productive way to model diverse expressions of a common source, while the technical challenges they identify have suggestive theological analogs (the hierarchical variant in which some traditions' decoders are more expressive; the tension between esoteric unity and exoteric diversity; the variation in contemplative ``encoder quality'' across traditions). To our knowledge, no prior work has systematically mapped the VAE framework onto the structure of religious epistemology.

\section{The Generative Ontology}
\label{sec:ontology}

We define the theological landscape as a probabilistic graphical model involving a latent variable $z$ and an observable variable $x$. A note on the formal apparatus is in order. The definitions, propositions, and assumptions below state precise structural claims within our model; they are not deductive theorems with supplied proofs. Where we invoke results from information theory (the Data Processing Inequality) or differential topology (transversality), those results hold rigorously in their native mathematical contexts; we apply them here as structural constraints that the analogy inherits from its mathematical source. The formal environments indicate precision of statement, not that the full proof obligations of pure mathematics have been discharged. The arguments of Sections~\ref{sec:exclusivism}--\ref{sec:perennialism} are abductive: they show that the perennialist configuration provides the best explanatory fit within the model, not that it follows as a mathematical theorem.

\begin{figure}[t]
\centering
\begin{tikzpicture}[
    box/.style={rectangle, draw=gray!40, line width=0.4pt, rounded corners, minimum width=2.6cm, minimum height=1.1cm, align=center, font=\small},
    arrow/.style={-{Stealth[length=3mm]}, thick}
]
    \definecolor{neutral}{RGB}{130,130,130}
    \definecolor{tradgen}{RGB}{70,140,80}

    % Latent space
    \node[box, fill=neutral!15] (z) {Latent Space $\mathcal{Z}$ \\ \textbf{The Absolute}};

    % Observable space
    \node[box, fill=tradgen!15, right=4cm of z] (x) {Observable Space $\mathcal{X}$ \\ \textbf{Exoteric Forms}};

    % Prior
    \node[above=1cm of z, align=center] (prior) {$p(z)$ \\ \footnotesize\textit{Undifferentiated Prior}};
    \draw[arrow, dashed, gray] (prior) -- (z);

    % Decoder (Revelation) - curved arrow on top
    \draw[arrow, bend left=25, tradgen!70!black, line width=1.2pt] (z.north east) to
        node[above, align=center, yshift=2pt] {$p_\theta(x|z)$ \\ \footnotesize\textbf{Revelation / Decoder}}
        (x.north west);

    % Encoder (Contemplation) - curved arrow on bottom
    \draw[arrow, bend left=25, tradgen!70!black, line width=1.2pt] (x.south west) to
        node[below, align=center, yshift=-2pt] {$q_\phi(z|x)$ \\ \footnotesize\textbf{Contemplation / Encoder}}
        (z.south east);

    % Parameter key below the diagram
    \coordinate (midpoint) at ($(z)!0.5!(x)$);
    \node[below=2.5cm of midpoint, draw=gray!40, line width=0.4pt, rounded corners, fill=gray!5, inner sep=8pt, font=\footnotesize] (params) {%
        \begin{tabular}{@{}ll@{\quad}ll@{}}
        \multicolumn{2}{c}{\textbf{\color{tradgen!70!black}Decoder $\theta$}} & \multicolumn{2}{c}{\textbf{\color{tradgen!70!black}Encoder $\phi$}} \\[3pt]
        $\theta_{\text{Christ}}$: & Incarnation, Liturgy & $\phi_{\text{Christ}}$: & Hesychasm, Lectio Divina \\
        $\theta_{\text{Islam}}$: & Prophecy, Sharia & $\phi_{\text{Islam}}$: & Dhikr, Salat \\
        $\theta_{\text{Vedanta}}$: & Upanishads, Brahman & $\phi_{\text{Vedanta}}$: & Jñāna Yoga, Dhyāna
        \end{tabular}%
    };

\end{tikzpicture}
\caption{The VAE-theology mapping (Definitions \ref{def:latent-absolute}--\ref{def:encoder}). Successful inference requires that the encoder $\phi$ be matched to the decoder $\theta$ whose outputs it receives.}
\label{fig:vae-theology}
\end{figure}
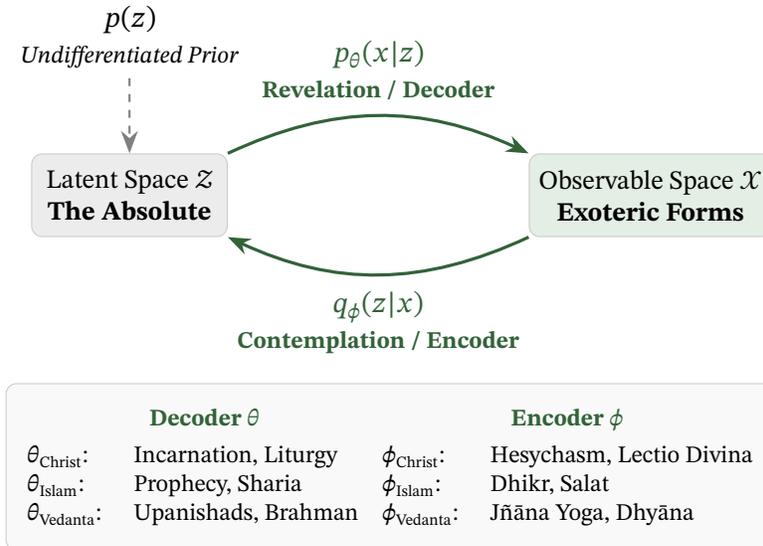

\begin{definition}[\textbf{The Latent Absolute, $z$}]
\label{def:latent-absolute}
    Let $z \in \mathcal{Z}$ be a low-dimensional latent vector representing the "Absolute," "Dao," or "Godhead." It follows a prior distribution $p(z)$, taken to be isotropic (i.e., $\mathcal{N}(0, I)$ in the standard VAE setting, representing maximal symmetry and the absence of privileged directions), modeling the undifferentiated, essential core of truth prior to cultural instantiation. The low dimensionality of $z$ reflects the theological principle of divine simplicity, originating in Plotinus's doctrine that the One is beyond all predication and multiplicity \cite[V.4.1--2]{plotinus_enneads}, later systematized by Aquinas \cite[I, q.~3]{aquinas}: the source is compact and unified, while its cultural manifestations are richly elaborated. Whether a single $\mathcal{Z}$ is shared across traditions or each tradition possesses its own latent space is precisely the question under investigation; Section~\ref{sec:configurations} introduces the competing configurations. The apophatic objection that the Absolute resists all formal characterization, and the Buddhist non-substantivist objection, are addressed in Section~\ref{sec:latent-absolute-status}.
\end{definition}

\begin{definition}[\textbf{The Exoteric Observable, $x$}]
    Let $x \in \mathcal{X}$ be the observable data of a religion, that is, the outputs of the tradition's generative process. This includes doctrinal propositions (e.g., the Nicene Creed, the Four Noble Truths), liturgical forms (the Divine Liturgy, the Salat), sacred texts (the Bible, the Quran, the Pali Canon), ethical codes (the Ten Commandments, the Vinaya), ceremonial practices, iconography, sacred architecture, and devotional music. The space $\mathcal{X}$ represents the total space of possible human cultural artifacts and behaviors. These observables are what the practitioner encounters and what the contemplative practices (the encoder, Definition~\ref{def:encoder}) take as input.
\end{definition}

\begin{definition}[\textbf{Revelation as Decoder, $p_\theta(x|z)$}]
    Revelation is the generative decoder function mapping the latent absolute to the observable world. It is parameterized by $\theta$, which represents the specific "covenant" or cultural-historical context of a tradition (e.g., Christianity, Buddhism). The decoder is not a practice performed by individuals but the tradition's historically accumulated generative process: the means by which transcendent truth has been \textit{expressed} in the observable forms described above.
    \begin{equation}
        x \sim p_\theta(x|z)
    \end{equation}
    Because $\theta$ varies by tradition, different religions generate different observables $x$. Under one configuration (Configuration~P, Section~\ref{sec:configurations}), these observables derive from a shared source $z$; under another (Configuration~E), from independent sources $z_i$.
\end{definition}

\begin{definition}[\textbf{Contemplation as Encoder, $q_\phi(z|x)$}]
\label{def:encoder}
    The encoder is the \textit{inverse} of the decoder: where $p_\theta$ maps from the Absolute to observable form, $q_\phi$ maps from observable form back toward the Absolute. Concretely, $q_\phi(z|x)$ represents the contemplative, mystical, and ascetic practices through which a practitioner attempts to recover the latent source $z$ from the tradition's observable data $x$. These include systematic disciplines such as the Buddhist \textit{jhānas} (progressive stages of meditative absorption), Teresa of Ávila's stages of the Interior Castle, the Sufi \textit{maqāmāt} (stations of the spiritual path), and the Kabbalistic ascent through the \textit{sefirot} (the ten divine emanations). Each constitutes a tradition-specific method, parameterized by $\phi$, for inferring the hidden structure that generated the tradition's observables.
    \begin{equation}
        z \sim q_\phi(z|x)
    \end{equation}
    The encoder $\phi$ is what the practitioner \textit{trains} through sustained contemplative engagement. It is not given by revelation but learned through disciplined practice within a specific tradition.
\end{definition}

The Traditionalist school's central structural claim is that every tradition possesses an exoteric (outer, public, legal-ritual) dimension and an esoteric (inner, initiatic, contemplative) dimension, and that the transcendent unity of religions holds at the esoteric level only \cite{schuon1948}. Our framework maps this distinction naturally: the exoteric dimension corresponds to the decoder $p_\theta$ and the observable space $\mathcal{X}$, while the esoteric dimension corresponds to the encoder $q_\phi$ and the latent space $\mathcal{Z}$. Not all traditions explicitly recognize this distinction. Protestant Christianity and mainstream Islamic scholarship, for example, reject formal esoteric hierarchies. Yet even these traditions contain contemplative currents such as the Quaker inner light, Hesychasm, and Sufi practice within Sunni orthodoxy that operationally distinguish between outward observance and inward realization. Our framework requires only this weaker structural claim: that the distinction between observable forms (decoder outputs) and contemplative inference (encoder operation) can be realized in practice.

\subsection{Three Competing Configurations}
\label{sec:configurations}

The definitions above leave open a crucial structural question: is the latent space $\mathcal{Z}$ shared across traditions, and if so, how do tradition-specific decoders relate to it? We identify three competing configurations of the generative model, each corresponding to a major position in the philosophy of religion:

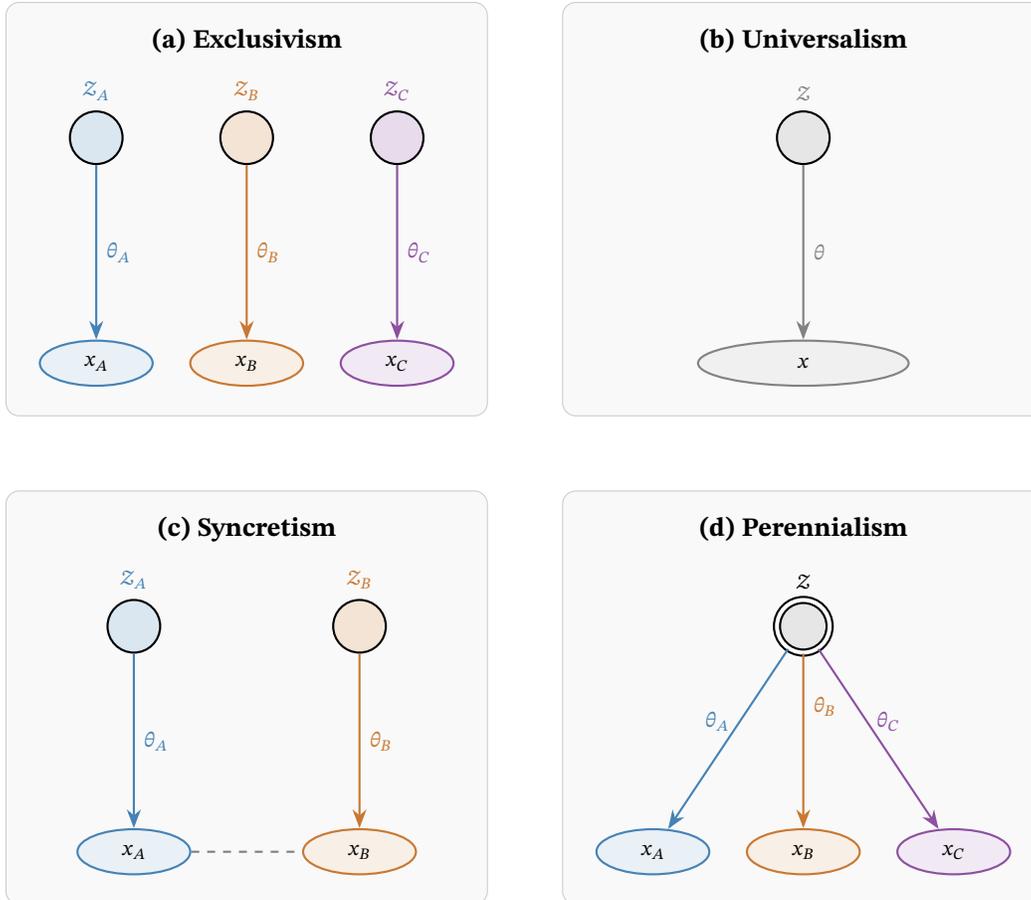
\begin{figure}[t]
\centering
\begin{tikzpicture}[
    zcirc/.style={circle, draw, thick, fill=#1!20, minimum size=7mm, inner sep=0pt},
    mshape/.style={ellipse, draw=#1, thick, fill=#1!12, minimum width=15mm, minimum height=6mm, inner sep=0pt},
    dec/.style={-{Stealth[length=2.5mm]}, thick, #1},
    plabel/.style={font=\small\bfseries},
    zlabel/.style={font=\scriptsize, #1},
    mlabel/.style={font=\scriptsize},
    tlabel/.style={font=\scriptsize},
    panelbox/.style={rounded corners=5pt, fill=gray!5, draw=gray!40, line width=0.4pt},
]
    \definecolor{tradA}{RGB}{70,130,180}
    \definecolor{tradB}{RGB}{200,120,50}
    \definecolor{tradC}{RGB}{140,80,160}
    \definecolor{neutral}{RGB}{130,130,130}

    % === Panel (a): Exclusivism (top-left) ===
    \draw[panelbox] (-0.2, 0.0) rectangle (6.2, 5.5);
    \node[plabel] at (3, 5.0) {(a) Exclusivism};

    \node[zcirc=tradA] (za1) at (1, 3.7) {};
    \node[zlabel=tradA, above] at (za1.north) {$\mathcal{Z}_A$};
    \node[mshape=tradA] (ma1) at (1, 0.7) {};
    \node[mlabel] at (ma1) {$x_A$};
    \draw[dec=tradA] (za1) -- node[right, tlabel] {$\theta_A$} (ma1);

    \node[zcirc=tradB] (zb1) at (3, 3.7) {};
    \node[zlabel=tradB, above] at (zb1.north) {$\mathcal{Z}_B$};
    \node[mshape=tradB] (mb1) at (3, 0.7) {};
    \node[mlabel] at (mb1) {$x_B$};
    \draw[dec=tradB] (zb1) -- node[right, tlabel] {$\theta_B$} (mb1);

    \node[zcirc=tradC] (zc1) at (5, 3.7) {};
    \node[zlabel=tradC, above] at (zc1.north) {$\mathcal{Z}_C$};
    \node[mshape=tradC] (mc1) at (5, 0.7) {};
    \node[mlabel] at (mc1) {$x_C$};
    \draw[dec=tradC] (zc1) -- node[right, tlabel] {$\theta_C$} (mc1);

    % === Panel (b): Universalism (top-right) ===
    \begin{scope}[shift={(7.4, 0)}]
        \draw[panelbox] (-0.2, 0.0) rectangle (6.2, 5.5);
        \node[plabel] at (3, 5.0) {(b) Universalism};

        \node[zcirc=neutral] (z2) at (3, 3.7) {};
        \node[zlabel=neutral, above] at (z2.north) {$\mathcal{Z}$};
        \node[mshape=neutral, minimum width=28mm] (m2) at (3, 0.7) {};
        \node[mlabel] at (m2) {$x$};
        \draw[dec=neutral] (z2) -- node[right, tlabel] {$\theta$} (m2);
    \end{scope}

    % === Panel (c): Syncretism (bottom-left) ===
    \begin{scope}[shift={(0, -6.5)}]
        \draw[panelbox] (-0.2, 0.0) rectangle (6.2, 5.5);
        \node[plabel] at (3, 5.0) {(c) Syncretism};

        \node[zcirc=tradA] (za3) at (1.5, 3.7) {};
        \node[zlabel=tradA, above] at (za3.north) {$\mathcal{Z}_A$};
        \node[mshape=tradA] (ma3) at (1.5, 0.7) {};
        \node[mlabel] at (ma3) {$x_A$};
        \draw[dec=tradA] (za3) -- node[right, tlabel] {$\theta_A$} (ma3);

        \node[zcirc=tradB] (zb3) at (4.5, 3.7) {};
        \node[zlabel=tradB, above] at (zb3.north) {$\mathcal{Z}_B$};
        \node[mshape=tradB] (mb3) at (4.5, 0.7) {};
        \node[mlabel] at (mb3) {$x_B$};
        \draw[dec=tradB] (zb3) -- node[right, tlabel] {$\theta_B$} (mb3);

        % Dashed line showing x-space combination
        \draw[thick, dashed, gray] (ma3) -- (mb3);
    \end{scope}

    % === Panel (d): Perennialism (bottom-right) ===
    \begin{scope}[shift={(7.4, -6.5)}]
        \draw[panelbox] (-0.2, 0.0) rectangle (6.2, 5.5);
        \node[plabel] at (3, 5.0) {(d) Perennialism};

        \node[zcirc=gray, double, double distance=1.5pt] (z4) at (3, 3.7) {};
        \node[zlabel=black, above] at (z4.north) {$\mathcal{Z}$};

        \node[mshape=tradA] (ma4) at (1, 0.7) {};
        \node[mlabel] at (ma4) {$x_A$};
        \node[mshape=tradB] (mb4) at (3, 0.7) {};
        \node[mlabel] at (mb4) {$x_B$};
        \node[mshape=tradC] (mc4) at (5, 0.7) {};
        \node[mlabel] at (mc4) {$x_C$};

        \draw[dec=tradA] (z4) -- node[left, tlabel, pos=0.4] {$\theta_A$} (ma4);
        \draw[dec=tradB] (z4) -- node[right, tlabel, pos=0.3] {$\theta_B$} (mb4);
        \draw[dec=tradC] (z4) -- node[right, tlabel, pos=0.4] {$\theta_C$} (mc4);
    \end{scope}

\end{tikzpicture}
\caption{Four configurations of the generative ontology. \textbf{(a)}~Exclusivism: each tradition possesses an independent latent space and decoder, with no shared structure (Definition~\ref{def:config-e}). \textbf{(b)}~Universalism: a single decoder maps one latent source to undifferentiated observables, erasing tradition-specific content (Definition~\ref{def:config-u}). \textbf{(c)}~Syncretism: distinct generative processes whose outputs are combined in the observable space (Section~\ref{sec:syncretism}). \textbf{(d)}~Perennialism: a shared latent space decoded through tradition-specific parameters, yielding distinct but structurally related observables (Definition~\ref{def:config-p}).}
\label{fig:configurations}
\end{figure}

\begin{definition}[\textbf{Configuration E: Exclusivism}]
\label{def:config-e}
    Each tradition $i$ possesses its own latent space $\mathcal{Z}_i$ and decoder $p_{\theta_i}(x|z_i)$. There is no shared latent structure: $\mathcal{Z}_A \cap \mathcal{Z}_B = \emptyset$ for distinct traditions $A$ and $B$.
\end{definition}

\begin{definition}[\textbf{Configuration U: Universalism}]
\label{def:config-u}
    All traditions share a single latent space $\mathcal{Z}$ and a single universal decoder $p_\theta(x|z)$. Tradition-specific differences are superficial variations in the observable space, not reflections of genuinely distinct generative processes.
\end{definition}

\begin{definition}[\textbf{Configuration P: Perennialism}]
\label{def:config-p}
    All traditions share a single latent space $\mathcal{Z}$, but each tradition possesses a distinct decoder $p_{\theta_i}(x|z)$ that maps the shared source to tradition-specific observables. The unity is in $\mathcal{Z}$; the diversity is in the decoders $\{\theta_i\}$.
\end{definition}

In addition to these three configurations, we evaluate a fourth approach that operates not on the latent space or decoders but on the observables directly:

\begin{definition}[\textbf{Syncretism}]
\label{def:syncretism}
    Syncretism combines or interpolates the observable outputs of distinct traditions in $\mathcal{X}$, without modifying the latent space $\mathcal{Z}$ or the decoder parameters $\theta$. Unlike Configurations E, U, and P, syncretism is not a claim about the generative structure of traditions, but an operation on decoder outputs: $x_{\text{sync}} = f(x_A, x_B)$ for some mixing function $f$.
\end{definition}

We now evaluate these configurations (Figure~\ref{fig:configurations}) through systematic elimination.

\section{The Failure of Exclusivism: Independent Latent Spaces}
\label{sec:exclusivism}

Configuration~E posits that each tradition operates with its own latent space $\mathcal{Z}_i$, sharing no structure with any other. This is the formal analog of strong exclusivism: one tradition accesses reality, and the rest are simply wrong. While internally consistent, Config~E faces a central explanatory problem: the extensive evidence for cross-traditional contemplative convergence.
\label{sec:convergence}

Comparative scholarship from James \cite{james1902} and Underhill \cite{underhill1911} through Stace \cite{stace1960} and Griffiths \cite{griffiths1986} documents striking convergences in the outputs of independently trained encoder functions $q_\phi$ across traditions that had no historical contact. The Buddhist \textit{jhānas} \cite{gethin1998}, Teresa of Ávila's stages of the Interior Castle \cite{teresa1577}, and the Sufi \textit{maqāmāt} \cite{schimmel1975} describe structurally similar trajectories through contemplative experience: progressive ego-dissolution, non-dual awareness, and ineffability, despite being developed as encoders for entirely different decoders. The parallel is strikingly specific: the progression through the four \textit{jhānas} into the formless attainments (infinite space, infinite consciousness, nothingness, neither-perception-nor-non-perception) mirrors the progressive ``emptying'' described in Teresa's Interior Castle, from discursive meditation through the prayer of quiet to spiritual marriage, and the Sufi stations leading to \textit{fanā} (annihilation in the divine). These independently developed encoder trajectories converge on structurally similar phenomenological endpoints, regardless of whether the tradition doctrinally affirms a substantive Absolute.

Beyond structural parallels in contemplative stages, practitioners who have trained encoder $q_{\phi_A}$ within one tradition have historically reported recognizing genuine contemplative attainment in practitioners who trained $q_{\phi_B}$ within another. The early Christian monastic tradition drew freely on Hellenistic spiritual exercises \cite{hadot1995}, and Thomas Merton's encounters with Asian contemplatives in the 1960s exemplify the modern recognition that practitioners of different traditions can identify shared depths of realization \cite{merton1973}. A well-trained encoder produces recognizable signatures of proximity to $z$; if the latent spaces were entirely separate, the outputs of $q_{\phi_A}$ and $q_{\phi_B}$ would share no common structure, and such cross-traditional recognition would be inexplicable.

Recent empirical work has advanced the evidence beyond textual comparison. Thomas Metzinger's concept of ``minimal phenomenal experience'' (MPE), the simplest possible form of conscious awareness corresponding to what contemplatives describe as ``pure consciousness'' or ``pure awareness,'' provides a scientifically operationalized framework for studying these convergences \cite{metzinger2020}. Gamma and Metzinger \cite{gamma2021} deployed a 92-item phenomenological questionnaire across 1,403 experienced meditators drawn from diverse traditions (Vipassanā, Zen, Śamatha, Transcendental Meditation, Mahāmudrā/Dzogchen, and others), finding a shared twelve-factor phenomenological structure that emerged consistently across traditions. The convergence was not limited to vague generalities: practitioners from traditions with incompatible doctrinal commitments reported the same specific cluster of features in statistically robust patterns. Metzinger's subsequent monograph \cite{metzinger2024}, drawing on over 500 experiential reports from meditators in 57 countries, confirms and extends these findings with a specificity and structural depth that resists explanation by observer bias or translation artifact alone.

In the language of our framework, these findings provide direct empirical evidence that independently trained encoders $q_{\phi_A}, q_{\phi_B}, \ldots, q_{\phi_N}$, operating on the outputs of distinct decoders, converge on a shared region of representational space. Given the reality of cross-traditional contemplative convergence, Config~E requires $N$ independent latent spaces for $N$ traditions and must treat the convergence as coincidence, an explanation that grows less parsimonious as the evidence accumulates. A shared latent space provides the most parsimonious explanation.

The convergence claim faces the well-known constructivist challenge of Katz \cite{katz1978}, who argues that contemplative experiences are wholly constructed by each tradition's conceptual framework, rendering cross-traditional comparison illegitimate. Our argument requires only the failure of \textit{strong} constructivism (the claim that no tradition-independent phenomenological residue exists); it is fully compatible with \textit{weak} constructivism, and indeed expects it, since the tradition-specificity of the decoder ensures that the form of experience differs across traditions while the shared latent space predicts convergence only at the deepest structural level. Metzinger's cross-traditional data (Section~\ref{sec:convergence}) provide empirical evidence against strong constructivism at a scale that earlier philosophical arguments could not achieve. We address the constructivist challenge and related objections in detail in Section~\ref{sec:constructivism}.

\section{The Failure of Syncretism: Non-Linear Decoding}
\label{sec:syncretism}

Exclusivism fails because it cannot explain convergence. Can one resolve this by combining traditions directly? Syncretism attempts exactly this: the combination or juxtaposition of elements from different traditions in the observable space $\mathcal{X}$. ``Christian Yoga,'' ``Zen Catholicism,'' New Age movements that freely mix Native American ritual, Eastern meditation, and Western occultism all combine the \textit{outputs} of different decoders rather than operating on the latent variable $z$ or on the decoder parameters $\theta$ themselves. Two critiques, developed below, show why this fails under all three configurations (E, U, and P). First, each tradition's observables lie on a distinct non-linear manifold in $\mathcal{X}$, and interpolation between manifolds lands off both. Second, and more fundamentally, because the decoders $p_\theta$ are non-linear, such interpolated points carry no guarantee of retaining the information about $z$ needed for any encoder to recover the Absolute.

\subsection{The Manifold Hypothesis}

\begin{assumption}[\textbf{The Manifold Hypothesis}]
\label{prop:manifold}
    Valid religious data points $x$ do not fill the space $\mathcal{X}$ uniformly but lie on low-dimensional, non-linear manifolds $\mathcal{M}_\theta \subset \mathcal{X}$.
\end{assumption}

This assumption requires justification. Religious traditions exhibit tight internal constraints, since doctrinal, liturgical, and ethical elements must cohere, so the space of practices satisfying these constraints is vastly smaller than the space of all possible practices. Moreover, traditions evolve through transmission chains that preserve structural relationships; arbitrary combinations would have been filtered out by centuries of development. Finally, if practices are generated from a latent source through a non-linear mapping $p_\theta(x|z)$, the image of this mapping necessarily lies on a manifold of dimension at most $\text{dim}(\mathcal{Z})$, typically much smaller than $\text{dim}(\mathcal{X})$. The first two arguments provide independent, model-external support for the manifold hypothesis; the third is conditional on the generative model itself.

\begin{claim}[\textbf{Manifold Disjointness}]
\label{prop:disjoint}
    For distinct tradition parameters $\theta_A \neq \theta_B$, the corresponding manifolds $\mathcal{M}_A$ and $\mathcal{M}_B$ are generically disjoint or intersect only at measure-zero sets.
\end{claim}

The motivation is geometric. In differential topology, the transversality theorem \cite{guillemin1974} establishes that two smooth maps from low-dimensional domains into a high-dimensional ambient space generically have disjoint images when the sum of their domain dimensions is less than the ambient dimension. We apply this as a structural constraint: insofar as religious traditions are generated by distinct non-linear mappings from a low-dimensional source into a vastly higher-dimensional observable space, their images are not expected to overlap. The theological interpretation is significant: there is no ``neutral ground'' in $x$-space where one might stand in two traditions simultaneously.

This model classifies \textit{naive} syncretism (interpolation) as failure. However, historical cases of apparent syncretism that succeed are better understood not as interpolation, but as the training of a \textit{new} decoder $\theta_{\text{new}}$ that generates a distinct manifold $\mathcal{M}_{\text{new}}$. Consider Sikhism, often characterized by outsiders as a Hindu--Islamic synthesis. The Sikh tradition itself insists on a distinct revelation from Guru Nanak \cite{mcleod1968}. In our terms, this amounts to a new decoder $\theta_{\text{new}}$ with its own access to $z$, not an interpolation of existing decoders. Success required the emergence of a new coherent manifold with its own structural logic, not a superposition of two existing ones. Such cases \textit{corroborate} the manifold critique: one cannot inhabit two existing manifolds simultaneously, and viable religious synthesis always takes the form of a new generative process. The possibility of such emergence is a signature prediction of the Perennialist (as distinct from the strict Traditionalist) view: because the Absolute can manifest new decoders spontaneously, the set $\{\theta_i\}$ is not fixed by a single primordial revelation but remains open to genuine religious innovation.

A distinct phenomenon from syncretism is the borrowing of encoder or decoder \textit{components} across traditions, analogous to transfer learning in machine learning. When a tradition adopts a contemplative technique from another (e.g., Christian \textit{lectio divina}, whose methods bear the imprint of Hellenistic reading practices \cite{leclercq1961}), this operates on the parameters $(\theta, \phi)$ themselves, not on their outputs. The result is a \textit{modified} tradition (an updated decoder or encoder) rather than a hybrid observable. We return to this distinction in Section~\ref{sec:discussion}.

\subsection{The Inferential Failure of Syncretism}

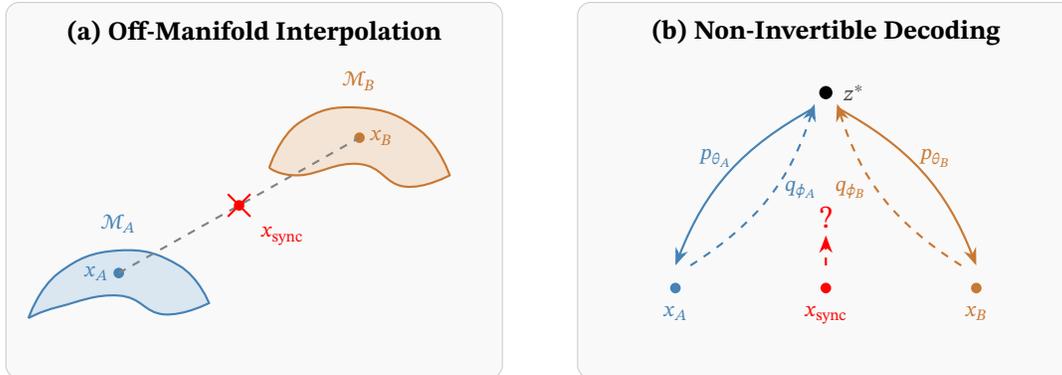
\begin{figure}[t]
\centering
\begin{tikzpicture}[
    plabel/.style={font=\small\bfseries},
    panelbox/.style={rounded corners=5pt, fill=gray!5, draw=gray!40, line width=0.4pt},
]
    \definecolor{manifoldA}{RGB}{70,130,180}
    \definecolor{manifoldB}{RGB}{200,120,50}
    \definecolor{interp}{RGB}{130,130,130}

    % === Panel (a): Manifold Interpolation ===
    \draw[panelbox] (-0.3, -0.8) rectangle (6.3, 4.2);
    \node[plabel] at (3, 3.8) {(a) Off-Manifold Interpolation};

    % Manifold A (curved surface - Christian)
    \draw[thick, manifoldA, fill=manifoldA!20]
        plot[smooth, tension=0.8] coordinates {(0,0) (0.4,0.6) (1.2,0.9) (2.0,0.7) (2.4,0.2)}
        -- plot[smooth, tension=0.8] coordinates {(2.4,0.2) (1.8,0.05) (1.2,0.3) (0.4,0.15) (0,0)};
    \node[manifoldA, font=\scriptsize, above] at (1.2,1.0) {$\mathcal{M}_A$};

    % Manifold B (curved surface - Hindu)
    \draw[thick, manifoldB, fill=manifoldB!20]
        plot[smooth, tension=0.8] coordinates {(3.2,2.0) (3.6,2.6) (4.4,2.8) (5.2,2.5) (5.6,1.9)}
        -- plot[smooth, tension=0.8] coordinates {(5.6,1.9) (5.0,1.75) (4.4,2.05) (3.6,1.9) (3.2,2.0)};
    \node[manifoldB, font=\scriptsize, above] at (4.4,2.9) {$\mathcal{M}_B$};

    % Points on manifolds
    \fill[manifoldA] (1.2,0.6) circle (2pt) node[font=\scriptsize, left] {$x_A$};
    \fill[manifoldB] (4.4,2.4) circle (2pt) node[font=\scriptsize, right] {$x_B$};

    % Interpolation line
    \draw[dashed, interp, thick] (1.2,0.6) -- (4.4,2.4);

    % Syncretic point (off manifold)
    \fill[red] (2.8,1.5) circle (2pt);
    \node[red, font=\scriptsize, below right] at (2.95,1.35) {$x_{\text{sync}}$};

    % X marks indicating "off manifold"
    \draw[red, thick] (2.65,1.65) -- (2.95,1.35);
    \draw[red, thick] (2.65,1.35) -- (2.95,1.65);

    % === Panel (b): Broken Round-Trip ===
    \begin{scope}[shift={(7.6, 0)}]
        \draw[panelbox] (-0.3, -0.8) rectangle (6.3, 4.2);
        \node[plabel] at (3, 3.8) {(b) Non-Invertible Decoding};

        % z* point at top
        \fill[black] (3.0, 3.0) circle (2.5pt);
        \node[font=\scriptsize\bfseries, black!70, right] at (3.1, 3.0) {$z^*$};

        % x_A point
        \fill[manifoldA] (1.0, 0.4) circle (2pt);
        \node[manifoldA, font=\scriptsize, below] at (1.0, 0.3) {$x_A$};

        % x_B point
        \fill[manifoldB] (5.0, 0.4) circle (2pt);
        \node[manifoldB, font=\scriptsize, below] at (5.0, 0.3) {$x_B$};

        % x_sync point
        \fill[red] (3.0, 0.4) circle (2pt);
        \node[red, font=\scriptsize, below] at (3.0, 0.3) {$x_{\text{sync}}$};

        % Decoder arrow z -> x_A (curved, non-linear)
        \draw[-{Stealth}, manifoldA, thick, bend right=20]
            (2.85, 2.8) to node[font=\scriptsize, left, pos=0.4] {$p_{\theta_A}$} (1.0, 0.7);

        % Encoder arrow x_A -> z (curved back)
        \draw[-{Stealth}, manifoldA, thick, dashed, bend right=20]
            (1.2, 0.7) to node[font=\scriptsize, right, pos=0.55] {$q_{\phi_A}$} (2.85, 2.85);

        % Decoder arrow z -> x_B (curved, non-linear)
        \draw[-{Stealth}, manifoldB, thick, bend left=20]
            (3.15, 2.8) to node[font=\scriptsize, right, pos=0.4] {$p_{\theta_B}$} (5.0, 0.7);

        % Encoder arrow x_B -> z (curved back)
        \draw[-{Stealth}, manifoldB, thick, dashed, bend left=20]
            (4.8, 0.7) to node[font=\scriptsize, left, pos=0.55] {$q_{\phi_B}$} (3.15, 2.85);

        % Broken arrow from x_sync (no encoder can invert)
        \draw[-{Stealth}, red, thick, dashed] (3.0, 0.7) -- (3.0, 1.1);
        \node[red, font=\large] at (3.0, 1.35) {?};

    \end{scope}

\end{tikzpicture}
\caption{The two failures of syncretism. \textbf{(a)}~Manifold interpolation: points $x_A$ and $x_B$ lie on their respective tradition manifolds, but their linear interpolation $x_{\text{sync}}$ lies off both manifolds in the ambient space $\mathcal{X}$. \textbf{(b)}~Non-invertible decoding: each tradition's observable is generated by a non-linear decoder ($p_{\theta_A}$, $p_{\theta_B}$) and can be inverted by a matched encoder ($q_{\phi_A}$, $q_{\phi_B}$) to recover $z^*$. The syncretic point $x_{\text{sync}}$ was generated by no decoder, and there is no guarantee that any encoder can recover $z^*$ from it.}
\label{fig:syncretism}
\end{figure}

The failure of syncretism is ultimately an inferential failure (Figure~\ref{fig:syncretism}). Let $x_A$ be a point on the Christian manifold $\mathcal{M}_A$ and $x_B$ be a point on the Hindu manifold $\mathcal{M}_B$. A syncretic approach attempts a linear interpolation:
\begin{equation}
    x_{\text{sync}} = \alpha x_A + (1-\alpha)x_B, \quad \alpha \in (0,1)
\end{equation}
Because the manifolds are non-linear and disjoint (Assumption~\ref{prop:manifold} and Claim~\ref{prop:disjoint}), the convex combination $x_{\text{sync}}$ lies \textit{off both manifolds} for all $\alpha \in (0,1)$. But the deeper problem is not merely geometric. Each valid observable $x \in \mathcal{M}_\theta$ was generated by a specific non-linear decoder $p_\theta(x|z)$ and therefore carries structured information about $z$ as refracted through $\theta$: an encoder $q_\phi$ can be trained to recover $z$ from such an $x$ because there exists a coherent generative process to invert. For $x_{\text{sync}}$, no such process exists. Because $p_\theta$ is non-linear, the blend of two decoders' outputs does not correspond to decoding through any single $\theta$, and there is no \textit{a priori} guarantee that the resulting point retains the information about $z$ needed to train an encoder capable of recovering the Absolute. The syncretic practitioner thus faces an unverifiable inference problem: they have constructed an observable with no generative model to invert.

To illustrate concretely: consider ``Christian Yoga'' as an attempted interpolation between Christian contemplative prayer ($x_A$) and Hindu \textit{asana} practice ($x_B$). Hesychastic prayer was developed to invert the Christian decoder: to recover $z$ from observables embedded in a framework of divine personhood, grace, and incarnational theology. Yogic practice was developed to invert the Hindu decoder: to recover $z$ from observables embedded in \textit{prana}, \textit{chakras}, and non-dual metaphysics. The interpolated practice (performing physical postures while reciting Christian prayers) was generated by neither decoder. Because $p_\theta$ is non-linear, the practitioner has no way to verify that this hybrid retains the information structure that either tradition's encoder requires to recover $z$.

In generative modeling, this is well-known: interpolating between images in pixel space produces incoherent superimpositions rather than meaningful hybrids, precisely because the non-linear decoder destroys the correspondence between pixel-space proximity and latent-space proximity \cite{bengio2013}. The syncretic error is analogous: performing surgery in the observable domain while ignoring the non-linear generative structure that gives observations their inferential power. A related failure, in which information about $z$ is destroyed not by interpolation but by reduction to common features, is analyzed in Section~\ref{sec:universalism}.

This analogy, however, invites an obvious objection: if meaningful interpolation in generative modeling operates in \textit{latent} space, why can we not interpolate in $\mathcal{Z}$ between traditions and thereby rescue syncretism? The answer is that every path through $z$ must exit through a \textit{specific} decoder. If one interpolates in $\mathcal{Z}$ and decodes through $\theta_A$, the output lies on $\mathcal{M}_A$; this is simply orthodox practice within tradition $A$, not syncretism. If one attempts to decode through both $\theta_A$ and $\theta_B$ simultaneously, say by averaging or superimposing outputs, the combination restates the $x$-space mixing problem at the output level, producing the same off-manifold artifacts. If one trains a new decoder $\theta_{\text{new}}$ to handle the interpolated latent representations, one arrives at manifold genesis, as argued above, not a hybrid inhabiting two existing manifolds. In no case does latent-space interpolation produce a valid point on both $\mathcal{M}_A$ and $\mathcal{M}_B$ simultaneously. The latent space confirms the Perennialist thesis: unity resides in $z$, but access to that unity always passes through a single decoder.

\section{The Failure of Universalism: Posterior Collapse}
\label{sec:universalism}

Exclusivism cannot explain convergence; syncretism cannot combine traditions without destroying their coherence. Can one instead extract what all traditions share? Configuration~U (Universalism) attempts this: it posits a shared latent space $\mathcal{Z}$ but a single universal decoder $\theta$, treating tradition-specific structure as cultural ``noise'' obscuring a universal ``signal.'' Examples include John Hick's pluralistic hypothesis \cite{hick1989} and certain forms of Unitarian Universalism.

\subsection{The Information-Theoretic Problem}

Mathematically, universalism is an attempt to reduce the full observable $x$ to a compressed representation $\tilde{x} = f(x)$ containing only features invariant across traditions. Consider what survives such reduction: perhaps some version of ``love your neighbor,'' ``practice compassion,'' and ``seek the transcendent.'' What is discarded? The Trinity, the Shahada, the Four Noble Truths, the Shema. These are precisely the doctrinal elements that distinguish traditions.

Let $f: \mathcal{X} \to \tilde{\mathcal{X}}$ be the universalist feature extractor. The key constraint is information-theoretic: processing data can only destroy information, never create it. By the Data Processing Inequality \cite{cover2006}, any deterministic function of $x$ cannot increase information about $z$:
\begin{equation}
    I(\tilde{x}; z) = I(f(x); z) \leq I(x; z)
\end{equation}
    with equality only if $f$ is a sufficient statistic for $z$ (i.e., a summary that retains all information relevant to inferring $z$). But the universalist projection is explicitly \textit{not} designed to preserve information about the latent; it is designed to preserve only features \textit{common} across different decoders. Since distinct decoders produce distinct observables from the same $z$, and since observables are generated as $x \sim p_\theta(x|z)$, the tradition-specific features of $x$ are not independent of $z$ but reflect $z$ as refracted through a particular $\theta$; discarding them discards information about $z$ itself, not merely about $\theta$. The projection is therefore not a sufficient statistic, yielding strict inequality:
\begin{equation}
    I(\tilde{x}; z) < I(x; z) \quad \text{(strict inequality)}
\end{equation}

\subsection{Posterior Collapse}

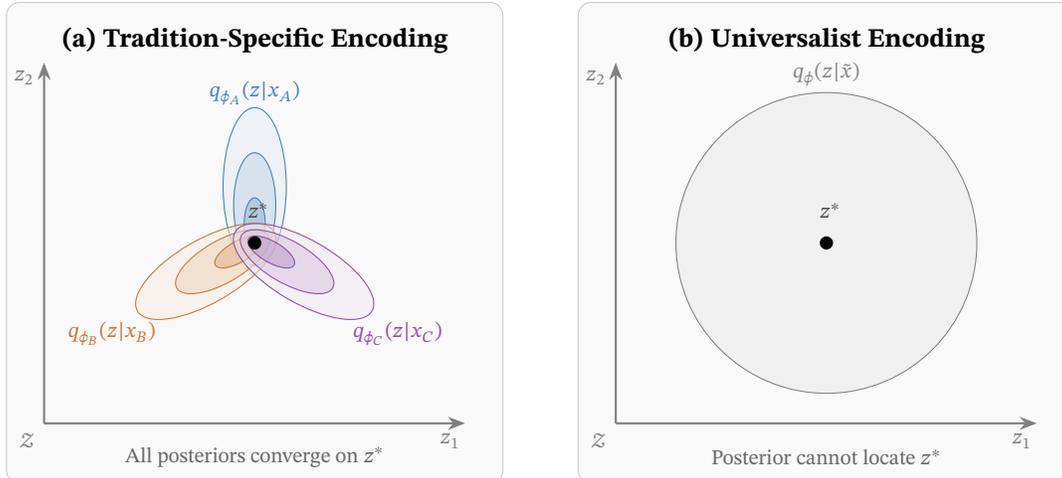
\begin{figure}[t]
\centering
\begin{tikzpicture}[
    plabel/.style={font=\small\bfseries},
    panelbox/.style={rounded corners=5pt, fill=gray!5, draw=gray!40, line width=0.4pt},
]
    \definecolor{tradA}{RGB}{70,130,180}
    \definecolor{tradB}{RGB}{200,120,50}
    \definecolor{tradC}{RGB}{140,80,160}
    \definecolor{neutral}{RGB}{130,130,130}

    % === Panel (a): Tradition-Specific Encoding ===
    \draw[panelbox] (-0.3, -0.6) rectangle (6.3, 5.8);
    \node[plabel] at (3, 5.3) {(a) Tradition-Specific Encoding};

    % 2D latent space axes
    \draw[-{Stealth}, thick, gray] (0.2, 0.2) -- (5.8, 0.2);
    \draw[-{Stealth}, thick, gray] (0.2, 0.2) -- (0.2, 5.0);
    \node[font=\scriptsize, gray, below] at (5.6, 0.2) {$z_1$};
    \node[font=\scriptsize, gray, left] at (0.2, 4.8) {$z_2$};
    \node[font=\scriptsize, gray, below left] at (0.2, 0.2) {$\mathcal{Z}$};

    % Posterior A - contour ellipses biased toward z*, from top
    \draw[tradA, thin, fill=tradA!15, fill opacity=0.5, rotate around={90:(3.0,3.35)}]
        (3.0, 3.35) ellipse (1.05 and 0.42);
    \draw[tradA, thin, fill=tradA!35, fill opacity=0.5, rotate around={90:(3.0,3.10)}]
        (3.0, 3.10) ellipse (0.7 and 0.28);
    \draw[tradA, thin, fill=tradA!60, fill opacity=0.5, rotate around={90:(3.0,2.85)}]
        (3.0, 2.85) ellipse (0.35 and 0.14);
    \node[font=\scriptsize, tradA] at (3.0, 4.6) {$q_{\phi_A}(z|x_A)$};

    % Posterior B - contour ellipses biased toward z*, from lower-left
    \draw[tradB, thin, fill=tradB!15, fill opacity=0.5, rotate around={30:(2.35,2.22)}]
        (2.35, 2.22) ellipse (1.05 and 0.42);
    \draw[tradB, thin, fill=tradB!35, fill opacity=0.5, rotate around={30:(2.57,2.35)}]
        (2.57, 2.35) ellipse (0.7 and 0.28);
    \draw[tradB, thin, fill=tradB!60, fill opacity=0.5, rotate around={30:(2.78,2.48)}]
        (2.78, 2.48) ellipse (0.35 and 0.14);
    \node[font=\scriptsize, tradB] at (1.1, 1.4) {$q_{\phi_B}(z|x_B)$};

    % Posterior C - contour ellipses biased toward z*, from lower-right
    \draw[tradC, thin, fill=tradC!15, fill opacity=0.5, rotate around={150:(3.65,2.22)}]
        (3.65, 2.22) ellipse (1.05 and 0.42);
    \draw[tradC, thin, fill=tradC!35, fill opacity=0.5, rotate around={150:(3.43,2.35)}]
        (3.43, 2.35) ellipse (0.7 and 0.28);
    \draw[tradC, thin, fill=tradC!60, fill opacity=0.5, rotate around={150:(3.22,2.48)}]
        (3.22, 2.48) ellipse (0.35 and 0.14);
    \node[font=\scriptsize, tradC] at (4.9, 1.4) {$q_{\phi_C}(z|x_C)$};

    % True z* marker (drawn last so it appears on top)
    \fill[black] (3.0, 2.6) circle (2.5pt);
    \node[font=\scriptsize\bfseries, black!70, above] at (3.05, 2.8) {$z^*$};

    % Annotation
    \node[font=\scriptsize, black!60, align=center] at (3.0, -0.25) {All posteriors converge on $z^*$};

    % === Panel (b): Universalist Encoding ===
    \begin{scope}[shift={(7.6, 0)}]
        \draw[panelbox] (-0.3, -0.6) rectangle (6.3, 5.8);
        \node[plabel] at (3, 5.3) {(b) Universalist Encoding};

        % 2D latent space axes
        \draw[-{Stealth}, thick, gray] (0.2, 0.2) -- (5.8, 0.2);
        \draw[-{Stealth}, thick, gray] (0.2, 0.2) -- (0.2, 5.0);
        \node[font=\scriptsize, gray, below] at (5.6, 0.2) {$z_1$};
        \node[font=\scriptsize, gray, left] at (0.2, 4.8) {$z_2$};
        \node[font=\scriptsize, gray, below left] at (0.2, 0.2) {$\mathcal{Z}$};

        % Collapsed posterior - large uniform circle (no peak, no structure)
        \draw[neutral, thin, fill=neutral!12]
            (3.0, 2.6) circle (2.0);
        \node[font=\scriptsize, neutral] at (3.0, 4.85) {$q_\phi(z|\tilde{x})$};

        % True z* marker (drawn last so it appears on top)
        \fill[black] (3.0, 2.6) circle (2.5pt);
        \node[font=\scriptsize\bfseries, black!70, above] at (3.05, 2.8) {$z^*$};

        % Annotation
        \node[font=\scriptsize, black!60, align=center] at (3.0, -0.25) {Posterior cannot locate $z^*$};
    \end{scope}

\end{tikzpicture}
\caption{Posterior collapse under universalism. \textbf{(a)}~Tradition-specific encoding: each tradition's full observable structure yields a tightly concentrated posterior (colored ellipses) that converges on the true latent value $z^*$, successfully recovering the Absolute through distinct paths. \textbf{(b)}~Universalist encoding: after reducing tradition-specific content to a common core $\tilde{x}$, the resulting posterior (gray) is too diffuse to locate $z^*$. The Data Processing Inequality guarantees this degradation: the universalist projection discards tradition-specific structure that carries information about $z$, producing a posterior that spreads across the latent space rather than concentrating on the Absolute.}
\label{fig:posterior-collapse}
\end{figure}

This information loss manifests as \textit{posterior collapse}, a phenomenon well-documented in the VAE literature \cite{bowman2016}. Posterior collapse occurs when the latent variable $z$ becomes independent of the data: the model learns to ignore specific structure and instead relies on a ``default'' output.

\begin{definition}[\textbf{Shallow Universalism as Posterior Collapse}]
    A generative model exhibits posterior collapse when $q_\phi(z|\tilde{x}) \approx p(z)$ for all $\tilde{x}$. In this regime, the latent variable carries no information about the specific input.
\end{definition}

In the standard VAE literature, posterior collapse is an \emph{optimization} pathology: the decoder becomes powerful enough to model the data without using $z$, or the KL penalty dominates early in training and pushes $q_\phi(z|x) \to p(z)$ before the reconstruction term can fight back. It is typically an accidental failure that can be mitigated through better training schedules (e.g., KL annealing \cite{bowman2016}) or architectural choices.

What universalism produces is an information-theoretic analog with the same symptom but a different cause. The Data Processing Inequality guarantees that the reduced input $\tilde{x}$ carries less information about $z$ than the full observable $x$, so any encoder trained on $\tilde{x}$ will produce a posterior that is, on average, closer to the prior and less able to disambiguate among competing accounts of $z$. No encoder, however well-optimized, can recover information that the pre-processing step has already destroyed. This makes the universalist failure distinct from standard posterior collapse in an important respect: the latter is an accidental optimization pathology that can be mitigated through better training schedules; the former is a \textit{necessary consequence} of the input bottleneck. Universalism creates an artificial bottleneck, the ``common core,'' that is too narrow to carry tradition-specific information, and the resulting inference is systematically degraded: a generic ``spirituality'' that drifts toward the undifferentiated prior. This information-theoretic failure parallels the syncretist failure (Section~\ref{sec:syncretism}): where syncretism destroys information about $z$ by blending the outputs of incompatible non-linear decoders, universalism destroys it by projecting away the tradition-specific structure through which $z$ is encoded. Both errors operate in $\mathcal{X}$ rather than $\mathcal{Z}$, and both produce observables from which the Absolute cannot be reliably recovered.

The posterior collapse interpretation explains why universalist spirituality often feels ``thin'' to committed practitioners. This is not mere prejudice but a structural consequence of information loss: a universalist representation retains only features common across all traditions, excluding the specific doctrinal, liturgical, and contemplative structures through which each tradition enables inference of $z$. Just as a compressed image that retains only features common to all photographs loses the details that make any particular photograph meaningful, the ``common core'' of all religions loses the particular structures (the Christian sacraments, the Buddhist eightfold path, the Islamic five pillars) through which practitioners actually engage with transcendent reality. The result is a spirituality formally compatible with all traditions but operationally equivalent to none: it preserves the vocabulary of the sacred while discarding the mechanisms of access.

\section{Perennialism: The Optimization of Orthodoxy}
\label{sec:perennialism}

Exclusivism cannot explain convergence. Syncretism destroys coherence. Universalism collapses into vacuity. What remains is Configuration~P: a shared latent space decoded through tradition-specific parameters. The apparent paradox of Perennialism, affirming transcendent unity while insisting on strict orthodoxy, dissolves when we model spiritual practice as the training of an encoder $q_\phi(z|x)$ that must be matched to its corresponding decoder. Cross-traditional convergence licenses the inference to a shared $z$; what remains is to show why accessing that shared source requires commitment to a single tradition's decoder.

The spiritual goal is to approximate the true posterior $p(z|x)$: given the tradition's observable forms $x$, experience the latent Divine $z$. The quality of this approximation can be measured by a single score that balances two competing pressures: fidelity to the tradition's specific forms, and approach toward the universal source. This score is the Evidence Lower Bound (ELBO), a tractable lower bound on the log-probability of the observed data that becomes tight as the approximation improves:

\begin{equation}
\label{eq:elbo}
    \mathcal{L}(\theta, \phi; x) = \underbrace{\mathbb{E}_{q_\phi(z|x)} [\log p_\theta(x|z)]}_{\text{Reconstruction: Orthodoxy}} - \underbrace{D_{KL}(q_\phi(z|x) || p(z))}_{\text{Regularization: Self-Transcendence}}
\end{equation}

\subsection{Reconstruction and the Necessity of Orthodoxy}

The first term of the ELBO (Figure~\ref{fig:vae-theology}), $\mathbb{E}_{q_\phi(z|x)} [\log p_\theta(x|z)]$, measures how well the inferred latent state explains the specific observable practice. This term is maximized when the encoder $q_\phi$ produces latent representations that, when decoded through $p_\theta$, accurately reconstruct the input.

\begin{claim}[\textbf{Encoder-Decoder Matching}]
\label{prop:matching}
    The ELBO (Equation~\ref{eq:elbo}) is maximized with respect to $\phi$ when the encoder parameters are functionally inverse to the decoder parameters $\theta$. That is, $\phi^* = \arg\max_\phi \mathcal{L}(\theta, \phi; x)$ satisfies $q_{\phi^*}(z|x) \approx p_\theta(z|x)$.
\end{claim}

The theological implication is immediate: the contemplative practice ($\phi$) must be the one designed to \textit{invert} the specific revelation ($\theta$). The practice of \textit{hesychia} (stillness) and the Jesus Prayer \cite{meyendorff1974}, for instance, constitutes an encoder ($\phi_{\text{Christ}}$) developed specifically to recover $z$ from the observable forms of Christian theology: the liturgy, the sacraments, the theology of \textit{theosis} (divinization) \cite{russell2004}. The Sufi practice of \textit{dhikr} (remembrance) \cite{schimmel1975} is calibrated to the Quranic revelation and the theology of \textit{tawhid} \cite{izutsu1964}, and Vedantic \textit{jñāna yoga} is calibrated to the Upanishadic teaching that \textit{Ātman} is identical with \textit{Brahman}. Using one tradition's encoder on another tradition's observables would be a category error: the encoder is calibrated to a different decoder's outputs. This reflects Schuon's insistence on commitment to a single spiritual form \cite{schuon1948}.

\subsection{Self-Transcendence and the Balance of the ELBO}

The second term of the ELBO, $-D_{KL}(q_\phi(z|x) \| p(z))$, penalizes the KL divergence (a measure of how far one probability distribution lies from another) between the practitioner's inferred distribution $q_\phi(z|x)$ and the true prior $p(z)$, the undifferentiated Absolute prior to any tradition-specific instantiation.

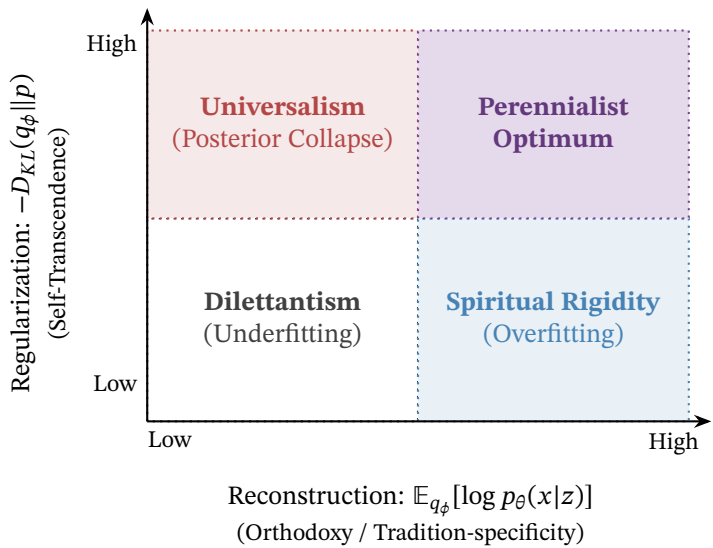
\begin{figure}[t]
\centering
\hspace*{-0.9cm}%
\begin{tikzpicture}[scale=1.0]
    % Define colors
    \definecolor{reconcolor}{RGB}{70,130,180}
    \definecolor{klcolor}{RGB}{180,70,70}
    \definecolor{optcolor}{RGB}{125,70,155}

    % Four quadrants divided at (3.6, 2.7), flush against axes

    % Bottom-left: Dilettantism (no fill)
    \draw[darkgray, thick, dotted] (0,0) rectangle (3.6,2.7);
    \node[darkgray, align=center, font=\small] at (1.8,1.35) {\textbf{Dilettantism} \\ (Underfitting)};

    % Top-left: Universalism
    \fill[klcolor!12] (0,2.7) rectangle (3.6,5.2);
    \draw[klcolor, thick, dotted] (0,2.7) rectangle (3.6,5.2);
    \node[klcolor, align=center, font=\small] at (1.8,3.95) {\textbf{Universalism} \\ (Posterior Collapse)};

    % Bottom-right: Spiritual Rigidity
    \fill[reconcolor!12] (3.6,0) rectangle (7.2,2.7);
    \draw[reconcolor, thick, dotted] (3.6,0) rectangle (7.2,2.7);
    \node[reconcolor, align=center, font=\small] at (5.4,1.35) {\textbf{Spiritual Rigidity} \\ (Overfitting)};

    % Top-right: Perennialist Optimum
    \fill[optcolor!20] (3.6,2.7) rectangle (7.2,5.2);
    \draw[optcolor, thick, dotted] (3.6,2.7) rectangle (7.2,5.2);
    \node[optcolor!80!black, align=center, font=\small\bfseries] at (5.4,3.95) {Perennialist \\ Optimum};

    % Axes (drawn last so they sit on top)
    \draw[-{Stealth}, thick] (0,0) -- (7.5,0);
    \draw[-{Stealth}, thick] (0,0) -- (0,5.5);

    % Axis labels
    \node[below, align=center] at (3.5,-0.7) {\small Reconstruction: $\mathbb{E}_{q_\phi}[\log p_\theta(x|z)]$ \\ \footnotesize (Orthodoxy / Tradition-specificity)};
    \node[above, rotate=90, align=center] at (-0.9,2.5) {\small Regularization: $-D_{KL}(q_\phi \| p)$ \\ \footnotesize (Self-Transcendence)};

    % Axis endpoint labels
    \node[below, font=\footnotesize] at (0.3,0) {Low};
    \node[below, font=\footnotesize] at (7,0) {High};
    \node[left, font=\footnotesize] at (0,0.5) {Low};
    \node[left, font=\footnotesize] at (0,5) {High};

\end{tikzpicture}
\caption{The ELBO optimization landscape for spiritual practice. The Perennialist optimum (upper right) balances reconstruction (commitment to tradition-specific forms) and regularization (self-transcendence). Dilettantism (lower left) represents underfitting: shallow engagement with neither traditional forms nor contemplative self-emptying. Universalism fails by abandoning specificity (left region). Spiritual rigidity fails by abandoning self-emptying (bottom region). (Note: the two axes are not independently controllable; both are functions of the shared parameters $(\theta, \phi)$. The landscape depicts the space of achievable ELBO decompositions, not independent degrees of freedom.)}
\label{fig:elbo}
\end{figure}

This term has a contemplative interpretation. Minimizing the KL divergence requires that the encoder's output, the practitioner's inferred posterior, approach the prior: the ``self'' that performs the inference must be progressively emptied until it reflects the undifferentiated source. Where the reconstruction term demands that the encoder $q_\phi$ be tradition-specific (matched to a particular decoder $p_\theta$), the regularization term demands that the \textit{result} of encoding approach a universal target.

\begin{definition}[\textbf{Self-Transcendence as KL Minimization}]
    Self-transcendence, known as \textit{kenosis} (Christianity), \textit{fan\=a} (Islam), or \textit{anatt\=a} (Buddhism), corresponds to minimizing the KL divergence. The practitioner's conditioned patterns of inference are dissolved until only the prior remains.
\end{definition}

This interpretation finds support across contemplative traditions. The archetype for this contemplative ascent is Plotinus's \textit{epistrophe} (return): the soul progressively strips away multiplicity and sensory particularity until it achieves union with the One \cite[VI.9.9--11]{plotinus_enneads}. Pseudo-Dionysius inherited this Neoplatonic framework and transmitted it to the Christian mystical tradition \cite{louth2007}, where it became the \textit{via negativa}. The \textit{via negativa} of Pseudo-Dionysius and Meister Eckhart emphasizes the stripping away of all concepts and images of God until one encounters the ``divine darkness,'' which is the prior before differentiation. Eckhart's paradoxical prayer to be rid of the conceptual God expresses KL minimization: even the tradition-specific concept of God must be surrendered to approach the unconditioned Absolute \cite{mcginn1981}. In Sufi terminology, \textit{fanā} (annihilation) describes an analogous process: the passing away of the ego-self in the divine presence, followed by \textit{baqā} (subsistence) in God \cite{schimmel1975}, as the practitioner's conditioned inference patterns ($q_\phi$) are dissolved into the unconditioned reality ($p(z)$). The Madhyamaka teaching of emptiness (\textit{śūnyatā}) systematically deconstructs the practitioner's reified concepts, revealing their lack of inherent existence \cite{garfield1995}. Strictly speaking, Madhyamaka does not posit a positive ground to ``return to''; \textit{śūnyatā} is a pure negation of inherent existence, mapping naturally to the KL minimization (the stripping away of reified constructs) without entailing a substantive $z$. We address this complexity in Section~\ref{sec:buddhist-non-substantivism}.

The ELBO (Equation~\ref{eq:elbo}; Figure~\ref{fig:elbo}) requires balancing both terms, and this trade-off mirrors a tension recognized across contemplative traditions between rigid attachment to form and formless dissolution. The parallel is structural rather than mechanistic: what the model captures is that fidelity to specific forms and approach to the universal are coupled constraints in a single objective, not that spiritual practice literally proceeds by gradient descent. On one extreme, maximizing only the reconstruction term without regularization leads to \textit{overfitting}: the practitioner develops a highly idiosyncratic inference network that perfectly ``explains'' the rituals but filters them through personal bias. The decoder works, but the latent representation is corrupted by the conditioned self. On the other extreme, maximizing only the KL term without reconstruction leads to \textit{posterior collapse}: the practitioner abandons the tradition entirely in pursuit of a contentless ``spirituality,'' the failure mode of universalism discussed in Section~\ref{sec:universalism}. The Perennialist insight is that both terms must be jointly optimized. One must commit fully to the tradition (maximizing reconstruction) while simultaneously undergoing the self-transcendence prescribed by that tradition (minimizing KL divergence). The decoder $p_\theta$ provides the structure (the observable forms that the encoder takes as input) while the encoder $q_\phi$ provides the mapping back toward the undifferentiated source. Neither alone suffices.

\section{Discussion and Limitations}
\label{sec:discussion}

The mapping between VAEs and religious epistemology has clear limitations. We address the principal objections, clarify the epistemic status of the argument, and explore several extensions.

\paragraph{The Question of Ground Truth.} In machine learning, we can verify whether a VAE successfully reconstructs data and produces meaningful latent representations. In theology, there is no analogous ``test set.'' The Absolute $z$ is, by definition, not directly accessible for empirical verification in the third-person sense. Our model thus describes the \textit{structure} of religious epistemology without adjudicating its \textit{truth}. That said, the contemplative traditions themselves insist that $z$ is accessible through direct first-person experience: the Christian mystic's \textit{unio mystica}, the Sufi's \textit{fanā}, the Buddhist's \textit{nibbāna}, the Hindu's \textit{mokṣa}. Countless practitioners across centuries and cultures have reported such experiences with remarkable consistency \cite{james1902, stace1960}. While these reports cannot serve as third-person verification, their convergence across independent traditions constitutes precisely the kind of evidence our model is designed to explain, and dismissing it entirely would beg the question against the framework's central premise.

\paragraph{Contemplative Scope.} Our framework models religious epistemology through the lens of contemplative practice: the encoder $q_\phi$ is realized through meditation, prayer, and ascetic discipline aimed at recovering a latent source. This focus is deliberate and reflects the Traditionalist school's own emphasis on the contemplative-mystical strand as the locus of transcendent unity. However, it entails a genuine limitation: traditions whose primary mode of engagement with the sacred is prophetic, legal, ethical, or communal (rabbinic Judaism centered on halakhic observance, Confucianism oriented toward ritual propriety and ethical cultivation, or indigenous traditions mediated through land, community, and ceremony) are not naturally modeled by an encoder-decoder architecture designed for contemplative inference. We do not claim that such traditions are deficient; we claim that the present framework is scoped to traditions with robust contemplative lineages, and that extending the model to prophetic, legal, and communal modes of religious life would require different formal apparatus.

\paragraph{The Apophatic Objection.}
\label{sec:latent-absolute-status} 
The mapping of the Absolute to $z$ invites an objection from the apophatic traditions: Pseudo-Dionysius, Meister Eckhart, Ibn Arabi, and Nāgārjuna all insist that the Absolute transcends \textit{all} categories, including mathematical structure. To assign $z$ a prior distribution might appear to domesticate what these traditions declare unknowable. But the framework already contains the apophatic insight. In variational inference, the ELBO (Section~\ref{sec:perennialism}) is a \textit{lower bound}, not an equality: there is always an irreducible gap between what the practitioner's inference achieves and the true structure of the source. Contemplative practice \textit{approximates} the Absolute without ever fully arriving at it. Eckhart's insistence that even the concept of God must be surrendered therefore expresses in traditional idiom what the formalism captures structurally.

\paragraph{Buddhist Non-Substantivism.} 
\label{sec:buddhist-non-substantivism}
A related objection arises from several Buddhist schools, most notably Madhyamaka and early Theravāda, which explicitly deny any substantive Absolute, making the assignment of $z$ appear to commit the category error identified by Katz \cite{katz1978}. We offer a two-pronged response. \textit{Phenomenologically}, our model does not require that a tradition doctrinally affirm $z$, but only that its encoder functions produce outputs structurally isomorphic to those of other traditions. The higher \textit{jhānas} and formless attainments culminating in cessation (\textit{nirodha-samāpatti}) produce phenomenological reports strikingly similar to descriptions of approaching the Absolute in theistic traditions \cite{griffiths1986}. The convergence of $q_\phi$ functions is the evidence, not doctrinal agreement about what $z$ is. \textit{Tradition-internally}, Buddhism is not monolithic on this question. Yogācāra posits \textit{ālaya-vijñāna} (store-consciousness) as a foundational substrate \cite{lusthaus2002}; Tathāgatagarbha traditions treat Buddha-nature as an innate Absolute \cite{king1991}; and Vajrayāna lineages identify clear light (\textit{prabhāsvara}) as the ultimate ground. The perennialist mapping is most natural for these schools. Where Madhyamaka and early Theravāda genuinely decline to posit $z$, the phenomenological convergence argument carries the burden. We acknowledge that this constitutes a genuine limitation: Madhyamaka and early Theravāda are hard cases, not edge cases, and the Yogācāra and Tathāgatagarbha doctrines on which we draw are themselves contested within Buddhist scholarship \cite{lusthaus2002}.

\paragraph{The Constructivist Challenge.}
\label{sec:constructivism}
The argument from convergence (Section~\ref{sec:convergence}) invites a powerful objection. Steven Katz \cite{katz1978} argues that there are no ``pure'' or ``unmediated'' mystical experiences: the contemplative practices of each tradition \textit{construct} the experiences they produce, so that a Christian mystic's vision of divine union and a Buddhist's experience of \textit{śūnyatā} are not two encounters with the same reality but two fundamentally different experiences shaped by incompatible conceptual frameworks. If Katz is right, the apparent convergences documented by James, Stace, and Griffiths are artifacts of decontextualized comparison, and the evidential basis for a shared latent space dissolves. The challenge comes in two versions. \textit{Strong constructivism} holds that contemplative experience is wholly constituted by the practitioner's prior conceptual framework, so that no tradition-independent phenomenological residue exists. \textit{Weak constructivism} holds that tradition shapes the interpretation and many features of contemplative experience, but does not fully determine it: there exist structural invariants in deep contemplative states that persist across traditions even when the practitioner's conceptual framework differs. Our argument requires only the failure of strong constructivism; it is fully compatible with weak constructivism, and indeed expects it: the tradition-specificity of the decoder $p_\theta$ ensures that the \textit{form} of contemplative experience differs across traditions, while the shared latent space $\mathcal{Z}$ predicts convergence only at the deepest structural level. Robert Forman's analysis of ``pure consciousness events'' (PCEs) provides the initial philosophical response to Katz. Forman \cite{forman1990} argues that contemplative traditions across cultures report a distinctive class of experience, wakeful awareness devoid of all intentional content, that cannot be the product of conceptual construction because it lacks the representational structure that construction requires. If an experience has no object, no image, and no conceptual content, there is nothing for the tradition's categories to have constructed. The existence of PCEs, if granted, constitutes a direct counterexample to strong constructivism and is what the perennialist framework predicts: an encoder output so thoroughly regularized (KL divergence approaching zero) that it approximates the undifferentiated prior $p(z)$ itself. The philosophical debate between Katz and Forman remained at an impasse for decades, in large part because both sides relied on textual interpretation and small numbers of case studies; Metzinger's empirical program (Section~\ref{sec:convergence}) has substantially shifted the balance by providing cross-traditional convergence data at a scale and specificity that strong constructivism struggles to accommodate.

\paragraph{The Naturalist Objection.} A sophisticated naturalist might grant the reality of convergence but deny its theological significance: brains are similar, so sufficiently advanced contemplative practices produce similar outputs regardless of tradition, ``convergence from below'' rather than ``convergence from above.'' Two responses. First, shared cognitive architecture readily explains \textit{generic} features of altered states (ego-dissolution, perceptual quieting), but it is less clear that it can account for the \textit{specific structural progressions} documented across traditions, the detailed stage-by-stage parallels between the jhānic sequence and the Teresian mansions, which involve not just similar endpoints but similar \textit{trajectories} through state space. Second, the ``convergence from below'' explanation is not incompatible with our framework. If shared neural architecture reliably produces convergent contemplative outputs, this architecture is itself a physical instantiation of the shared latent structure our model posits. The question of whether $\mathcal{Z}$ is ``transcendent'' or ``neural'' is metaphysical; the framework captures the \textit{structural fact} of a shared source underlying diverse expressions. But the two readings are not equivalent: if the shared structure is merely neural, the framework describes an interesting fact about human cognition but carries no theological weight; if it reflects a genuinely transcendent reality, the framework captures something about the relationship between the human and the divine. The Perennialist tradition maintains the latter. The formal model is compatible with it but does not settle the question.

\paragraph{Family Resemblance.} A final objection draws on Wittgenstein's concept of family resemblance: perhaps traditions share overlapping partial similarities without any single common essence, just as ``games'' share criss-crossing features with no defining property. On this view, no latent variable $z$ is needed; the similarities are real but unsystematic. The empirical evidence, however, tells against this prediction. Family resemblance predicts that different pairs of traditions will share \textit{different} features, with no single cluster present across all; Metzinger's data show the opposite: a specific twelve-factor phenomenological structure emerging consistently across traditions with incompatible doctrinal commitments. This systematic, cross-traditional convergence on a shared deep structure is the empirical signature of a common source (Configuration~P), not of mere family resemblance.

\paragraph{Epistemic Status.} We wish to be explicit about the logical character of our central claim. The argument for Configuration~P is \textit{abductive}, an inference to the best explanation, rather than deductive. We have not proven that a shared latent space exists; we have argued that, among the configurations considered, Config~P provides the most parsimonious and explanatorily adequate account of the observable structure of religious traditions. A natural circularity objection arises: does modeling religions as generative processes from a latent variable already presuppose perennialism? It does not. The generative framework is neutral among the configurations: Config~E uses separate latent spaces, Config~U uses a single decoder, and Config~P uses a shared space with distinct decoders. All three are expressible within the same formalism, and the exclusivist who denies any shared transcendent structure can reject Config~P without rejecting the modeling language itself. The perennialist conclusion is argued for through elimination, not built into the framework.

\paragraph{Testable Hypotheses.} The argument rests on two empirical premises: (a)~cross-traditional mystical convergence is genuine, and (b)~religious traditions possess sufficient internal coherence to be modeled as manifolds. Both are substantive and contestable; if either is rejected, the corresponding arguments lose their force. That said, the framework generates testable predictions: that practitioners who abandon tradition-specific practice for generic ``spirituality'' should report diminished contemplative depth, that conversion between traditions with structurally similar contemplative methods should proceed more readily than between distant ones, and that dimensionality reduction applied to corpora of contemplative texts should reveal shared low-dimensional structure. Metzinger's MPE data \cite{gamma2021, metzinger2024} already constitute a partial empirical test.

\paragraph{The Nature of the Prior.} The shared prior $p(z)$, a common Absolute underlying all traditions, is not assumed as a premise but argued for abductively through our comparison of Configurations~E, U, and~P (Sections~\ref{sec:exclusivism}--\ref{sec:perennialism}). An exclusivist might still argue that different traditions have different priors, or that only one tradition's prior corresponds to reality. Our framework is compatible with this view: simply set the ``true'' $p(z)$ to one tradition's conception, with others as approximations. Indeed, a significant strand of Perennialist thought takes precisely this position. Valentin Tomberg, the Christian Hermeticist whose \textit{Meditations on the Tarot} \cite{tomberg1985} synthesizes Catholic theology with Hermetic symbolism, maintained that the Catholic tradition expresses the fullest articulation of $p(z)$, while other traditions capture genuine but partial aspects of the same reality. On this reading, the decoders $\{\theta_i\}$ are not equally faithful to $z$: one tradition's decoder may produce observables that more completely reflect the latent structure, while others offer valid but less complete mappings. Configuration~P accommodates this hierarchical variant without modification; it requires only that the latent space be shared, not that all decoders be equally expressive. The abductive case for a shared prior rests on the explanatory advantages catalogued above; it does not foreclose either the egalitarian or hierarchical reading of the relationship among decoders.

\paragraph{Religious Conversion.} A significant challenge for our model is the phenomenon of religious conversion. If one has trained an encoder $\phi_A$ matched to decoder $\theta_A$, how can conversion to tradition $B$ be possible? On one reading, conversion involves abandoning the trained encoder $\phi_A$ and beginning training anew with $\phi_B$, a discontinuous parameter re-initialization rather than a smooth interpolation between traditions. This matches phenomenological accounts of conversion as ``dying and being reborn.'' On another reading, if the latent Absolute $z$ is truly shared across traditions, then a well-trained encoder $\phi_A$ may have learned representations of $z$ that partially transfer to tradition $B$. The convert does not start from zero but from an initialization informed by prior contemplative development. This would explain why mystics sometimes report recognizing truth in other traditions, and why conversion between traditions with similar contemplative structures may proceed more readily than conversion between distant ones.

\paragraph{Interfaith Dialogue.} If manifolds are disjoint and encoders must match decoders, what role remains for interfaith dialogue? The framework suggests that productive dialogue operates not in $x$-space, where traditions are incommensurable, but in $z$-space, where practitioners can recognize that others are engaged in the same project of latent variable inference, even if using different architectures. Such mutual recognition grounds respect without requiring synthesis. Dialogue may also help practitioners clarify the boundaries of their own manifolds: understanding what their tradition is \textit{not} can sharpen understanding of what it \textit{is}. Furthermore, comparing encoder architectures across traditions, noting where independently developed contemplative methods converge in their inferential structure, can yield indirect information about the geometry of $z$ itself. If multiple encoders, trained on different observables, produce similar posterior distributions, this constrains the latent space in ways that no single tradition's encoder could achieve alone. Interfaith dialogue thus serves not only ethical and social purposes but a genuinely epistemic one: the comparative study of contemplative methods can sharpen each tradition's inference of the Absolute.

\section{Conclusion}
\label{sec:conclusion}

By formalizing three competing configurations of the relationship between religious traditions and transcendent reality, and evaluating them through the lens of generative modeling, we have provided an abductive case for the Perennialist position. Orthodoxy, in this account, is not a rigid adherence to arbitrary rules but the preservation of the non-linear manifold structure required for valid inference. Our analysis yields four principal results:

\begin{enumerate}
    \item \textbf{Syncretism fails inferentially under all configurations.} Combining the outputs of distinct non-linear decoders in observable space produces artifacts that carry no guarantee of retaining the information about $z$ needed for encoder-based recovery of the Absolute. Latent-space interpolation either reduces to orthodox practice within a single tradition (when decoded through one $\theta$) or to tradition genesis (when a new decoder is trained), and in no case yields a valid hybrid inhabiting two existing manifolds (Section~\ref{sec:syncretism}).

    \item \textbf{Exclusivism cannot account for cross-traditional convergence.} Config~E, which posits independent latent spaces for each tradition, cannot parsimoniously explain the well-documented convergences in contemplative experience across traditions that developed in isolation (Section~\ref{sec:exclusivism}).

    \item \textbf{Universalism suffers from posterior collapse.} Config~U, which reduces traditions to a common core, discards the tradition-specific structure necessary for latent variable inference, resulting in an informationally impoverished representation that cannot recover the Absolute (Section~\ref{sec:universalism}).

    \item \textbf{Perennialism provides the best explanatory fit.} Configuration~P, a shared latent space with tradition-specific decoders, accounts for both convergence and diversity, and supplies the richest formal account of why committed orthodox practice maximizes the ELBO (Section~\ref{sec:perennialism}).
\end{enumerate}

These results carry implications for both religious practice and interfaith dialogue. For the practitioner, they provide formal justification for the traditional insistence on committed, integral practice within a single tradition. For the scholar of religion, they offer a vocabulary for discussing religious diversity that honors both the reality of cross-traditional convergence and the necessity of tradition-specific depth.

We do not claim to have proven the existence of a transcendent Absolute, nor the truth of any particular tradition. What this framework provides is a precise language, grounded in the mathematics of generative models, for articulating what the Perennialist tradition has long maintained on metaphysical grounds: that the diversity of religious forms is not evidence against transcendent unity, but the necessary consequence of a single source refracted through distinct generative processes, each requiring its own mode of contemplative return.

The ultimate test of this framework lies not in its formal elegance but in its alignment with contemplative experience. The traditions themselves have always insisted that their truth can only be verified through practice. Our formalization suggests why this must be so: the latent variable is accessible only through the encoder, and the encoder can only be trained through committed engagement with the tradition's data. Theory points beyond itself to practice; the map is not the territory; the model is not the Absolute.

\bibliography{references}

\end{document}